\documentclass[journal]{IEEEtran}

\usepackage{booktabs}
\usepackage{graphicx}
\usepackage{xspace}
\usepackage{url}

\newcommand{\ocinative}{\texttt{oci+native://}\xspace}
\newcommand{\ocifetch}{\texttt{oci+fetch://}\xspace}
\newcommand{\ociuri}{\texttt{oci://}\xspace}
\newcommand{\storageuri}{\texttt{storageUri}\xspace}

\newcommand{\prnative}{PR~\#5558\xspace}
\newcommand{\prfetch}{PR~\#5739\xspace}
\newcommand{\prcia}{PR~\#5770\xspace}
\newcommand{\prcib}{PR~\#5784\xspace}
\newcommand{\rfcint}{RFC~\#5789\xspace}

\begin{document}

\title{Cold-Start Model Delivery in Kubernetes Inference Serving:
An Empirical Study of OCI-Based Distribution and Its Integrity}

\author{Georgii~Kliukovkin%
\thanks{G. Kliukovkin is an independent researcher and KServe contributor
(e-mail: kliukovkin@gmail.com). Corresponding author: G.~Kliukovkin.}}

\markboth{Submitted to IEEE Access}{Kliukovkin: Cold-Start Model Delivery
in Kubernetes Inference Serving}

\maketitle

\begin{abstract}
The startup latency of a model-serving pod on Kubernetes is dominated by
one step: delivering the model weights. As models grow from megabytes to
the hundred-gigabyte weights of large language models, cold-start delivery
time governs the economics of autoscaling and scale-to-zero---every replica
added under load pays it before serving its first request. Yet the dominant
delivery mechanisms remain ad-hoc downloads from object storage, with none
of the pull caching, digest addressing, or verification that Kubernetes
provides for container images. We analyze the model delivery paths
available to a Kubernetes serving platform along two axes: which component
pulls the artifact, and whether any admission-time verifier can bind the
deployed reference to the bytes that arrive. We validate the analysis
upstream in KServe, a widely deployed Cloud Native Computing Foundation (CNCF)
model-serving platform, by implementing the two missing delivery paths: an \ocinative scheme that
mounts model images through Kubernetes image volumes (KEP-4639), merged
upstream, and an \ocifetch scheme that pulls OCI artifacts inside the
platform's storage initializer for clusters without image-volume support,
under review. We then report, to our knowledge, the first systematic
controlled comparison of model delivery paths within a Kubernetes serving
platform---modelcar sidecars, native image volumes, and object-storage
download; \ocifetch, still under upstream review, shares the
initializer data path and is left to future measurement---on artifacts
sized to the fp16
weights of 1B-, 7B-, and 70B-parameter models (2--140\,GB), measuring
cold- and warm-start time-to-first-prediction, per-phase latency, and node
storage overhead. Node-cached OCI delivery makes warm replica addition
size-independent---11.7\,s for a 70B-class artifact versus 40.7 minutes of
object-storage re-download, a $208\times$ difference---while the first cold
pull costs up to $2\times$ a plain download, a trade-off we localize to
containerd's blob-write-then-unpack double pass. Efficient delivery still leaves a trust gap: for
models that remain loose objects on \texttt{s3://}, \texttt{gs://}, or
\texttt{hf://} URIs, no admission-time verifier ever observes the bytes.
We therefore complement the distribution work with a serving-time
integrity design proposed to the KServe community---digest pinning and
Open Source Security Foundation (OpenSSF) model-signing enforcement in
the storage initializer, the last platform-owned component on the data
path---and show that hashing the artifact while it downloads adds under
0.1\% to delivery time, whereas a post-download verification pass adds
up to 53\%.
\end{abstract}


\section{Introduction}
\label{sec:intro}

Serving a machine-learning model on Kubernetes begins with moving its
weights. Before an inference pod can answer its first request, the model
artifact---megabytes for classic predictive models, up to hundreds of
gigabytes of fp16 weights for large language models---must travel from
wherever it is stored to the node that will serve it. That delivery step
dominates cold-start latency, and cold-start latency governs the economics
of GPU serving: how aggressively a fleet can scale to zero, how fast it can
scale out under load, and how much expensive accelerator time is spent
waiting for bytes rather than serving requests. Every replica added by an
autoscaler pays the delivery cost in full before it contributes a single
prediction.

Yet in the dominant deployment pattern the model travels the least-engineered
path in the stack. The model is \emph{not} in the container
image: a serving platform such as KServe~\cite{kserve} materializes an
inference pod whose init container---the \emph{storage
initializer}---downloads the model at startup from an object store or model
hub (\texttt{s3://}, \texttt{gs://}, \texttt{hf://}, \texttt{https://}) and
hands the bytes to the model server. The split is operationally sensible:
models are re-trained on a different cadence than serving code, they are
large, and they are produced by different teams. But the consequences
compound at LLM scale. The artifact is re-downloaded on every pod start
because nothing on the node caches it; it is copied at least once more on
its way to the server; and every added replica multiplies both costs.
Container images solved these exact problems years ago---kubelet-managed
pulls, node-level caching, content-addressed storage---and that machinery
has recently become applicable to models: Kubernetes image volumes
(KEP-4639~\cite{kep4639}), stable since v1.36, let a pod mount an arbitrary
OCI image as a read-only volume, and packaging efforts such as CNCF
ModelPack~\cite{modelpack} and KitOps~\cite{kitops} standardize how models
are laid out inside OCI images. What has been missing is the
serving-platform half: making the standard model-serving control plane
treat OCI-packaged models as a first-class storage source.

Delivery efficiency, however, is only half of the story: the container
ecosystem also gives images a trust apparatus that models entirely lack.
A container image is addressed by a
cryptographic digest, its provenance can be attested with in-toto
metadata~\cite{intoto} and signed with Sigstore~\cite{sigstore}, and
cluster operators enforce those signatures \emph{before the pod is
scheduled} with admission controllers such as Kyverno~\cite{kyverno} or the
sigstore policy-controller~\cite{policycontroller}; frameworks like
SLSA~\cite{slsa} grade the rigor of the whole pipeline. The model's
download path has none of this: it is addressed by a mutable reference,
verified against nothing, and performed after every admission decision has
already been made. A compromised bucket, a re-pointed model-hub revision,
or a man-in-the-middle on an \texttt{https://} pull all end the same way:
the pod starts and serves unverified weights. Any redesign of model
delivery should therefore answer for both properties at once---how fast the
bytes arrive, and how they can be trusted.

This paper describes work on both halves of that problem, carried out
upstream in KServe within a community roadmap to harmonize OCI model support
(issue \#4083). We make four contributions:

\begin{itemize}
\item \textbf{An analysis of the model delivery design space.} We
characterize the delivery paths available to a Kubernetes serving platform
along two axes---which component pulls the artifact (which determines
caching and copy behavior, hence cold-start cost), and whether an
admission-time verifier can bind the deployed reference to the arriving
bytes---and derive a placement rule for integrity enforcement
(Section~\ref{sec:distribution}, Table~\ref{tab:design-space}). The axes are
platform-generic; KServe is our implementation and measurement vehicle.
\item \textbf{Upstream validation: two new delivery paths.} We designed and
implemented the two previously missing points in that design space: the
\ocinative scheme, which materializes a model image as a Kubernetes image
volume mounted directly into the inference container---no sidecar, no
init-container copy, kubelet-managed pull and cache---merged
upstream (\prnative), with compatibility handling across the
image-volume API maturity window; and the \ocifetch scheme (\prfetch, under
community review), which pulls OCI artifacts inside the storage initializer
itself, runtime-agnostically (Section~\ref{sec:distribution}).
\item \textbf{Measurement.} To our knowledge we report the first systematic,
controlled comparison of model delivery paths within a Kubernetes serving
platform; no such data has, to our knowledge, been published for any of
these paths. The comparison covers modelcar, \ocinative, and \texttt{s3://} download
(three of the four design-space paths; \ocifetch is not yet merged and
is left to future measurement), on artifacts sized to the fp16 weights
of 1B-, 7B-, and 70B-parameter models (2, 14, and 140\,GB), cold and
warm, with per-phase startup latency, kubelet pull time, and node
storage overhead (Section~\ref{sec:eval}).
\item \textbf{Serving-time integrity for non-OCI URIs.} Applying the
placement rule to the models that remain loose objects on blob
storage---where no admission-time component ever observes the bytes---we
present a verification design proposed to the KServe community
(\rfcint): digest pinning and OpenSSF model-signing~\cite{modelsigning}
enforcement in the storage initializer, the integrity counterpart of the
platform's merged confidential-serving support, with its key feasibility
question answered by a streaming-hash microbenchmark
(Section~\ref{sec:integrity}, Section~\ref{sec:eval}).
\end{itemize}

We conclude with lessons for evolving the supply-chain posture of a large
CNCF project incrementally (Section~\ref{sec:discussion}). The benchmark harness
and all measurement data are open source at \url{https://github.com/kliukovkin/oci-model-delivery-bench}.

\section{Background and Related Work}
\label{sec:background}

\subsection{KServe and the model delivery problem}
\label{sec:kserve-bg}

KServe is a widely deployed CNCF (incubating) model-inference platform for
Kubernetes: an \texttt{InferenceService} custom resource describes a model
to serve, and controllers materialize the serving stack around it. The part
of that stack relevant here is model delivery. Each \texttt{InferenceService}
names its model with a \storageuri; a webhook injects a \emph{storage
initializer} init container that resolves the URI scheme (\texttt{s3://},
\texttt{gs://}, \texttt{hf://}, \texttt{https://}, \ldots), downloads the
artifact to a shared \texttt{emptyDir} volume, and exits before the model
server starts. The storage initializer is thus the single platform-owned
component that observes every downloaded model byte---a property that both
the confidential-serving work discussed below and our integrity proposal
build on.

KServe's first OCI-based alternative is the \emph{modelcar} pattern: the
model is packaged into a container image under \texttt{/models/}, and the
platform runs that image as a sidecar container in the inference pod, sharing
its filesystem with the model server. Modelcar demonstrated the appeal of OCI
distribution for models---registry infrastructure, layer dedup, image
pull caching---but pays for it with a permanently resident sidecar per pod
and container-runtime plumbing that predates purpose-built kernel and
runtime support for artifact mounting. A community roadmap (issue \#4083)
set out to harmonize modelcar with the newer alternatives that this paper
implements.

\subsection{OCI artifacts and Kubernetes image volumes}

The OCI image and distribution specifications have evolved from ``container
image format'' into a general content-addressed artifact ecosystem, and
Kubernetes has followed: KEP-4639~\cite{kep4639} introduces the
\texttt{image} volume source, letting a pod mount an arbitrary OCI image as
a read-only volume. The feature graduated deliberately: alpha in v1.31, beta
(still feature-gated) in v1.33, enabled by default in v1.35, and stable in
v1.36; containerd supports it from v2.1 (v2.2 for \texttt{subPath} mounts)
and CRI-O from v1.31. Image volumes carry exactly the properties the
model-delivery path lacks: kubelet-managed pull with node-level caching,
digest-addressable references, and---critically for supply-chain
enforcement---\emph{presence in the pod specification}, where admission
controllers can in principle verify the reference before scheduling
(Section~\ref{sec:relwork-security} discusses how far today's tooling actually
goes). Section~\ref{sec:distribution} describes how \ocinative maps model delivery
onto this primitive and what compatibility handling that multi-release API
maturity window required. A rich systems literature optimizes container
\emph{image distribution} itself---lazy pulling, dedup\-li\-ca\-tion, and
registry design, from Slacker~\cite{slacker} onward; it measures rootfs
provisioning for containers, not the delivery of model artifacts whose
paths differ in admission visibility, which is the axis this paper is
about.

\subsection{Model packaging and supply-chain security}
\label{sec:relwork-security}

\paragraph{Packaging.} CNCF ModelPack~\cite{modelpack} specifies how models,
tokenizers, and metadata are laid out as OCI artifacts; KitOps~\cite{kitops}
provides tooling that packages models as signed ``ModelKits,'' and has
recently been positioned as a route to verified inference on KServe
itself~\cite{kitops-talk}. These efforts answer \emph{how a model becomes a
(signed) OCI artifact}, and their route to verification requires
repackaging. They are complementary to this paper, which answers how the
serving platform \emph{delivers} such artifacts (Section~\ref{sec:distribution})
and how it can \emph{verify} the majority of models that are not repackaged
at all (Section~\ref{sec:integrity}).

\paragraph{Signing and verification.} in-toto~\cite{intoto} and
Sigstore~\cite{sigstore} established general software-artifact provenance
and signing; SLSA~\cite{slsa} frames them into pipeline maturity levels.
For ML specifically, the OpenSSF Model Signing (OMS)
specification~\cite{modelsigning} defines a sigstore-bundle-based signature
format over file-level hashes of a model directory, with keyless and
key-based flows; producer-side adoption is real---NVIDIA signs its NGC
catalog models with OMS~\cite{ngc-signing}---but serving-side enforcement
on the download path remains manual. The closest system to our proposal is
the sigstore \emph{model-validation-operator}~\cite{mvo}, an admission
webhook that injects an OMS-verification init container into labeled pods,
shipped downstream in Red Hat Trusted Artifact Signer~\cite{rhtas13}. It is
deliberately platform-agnostic, and that is precisely its limit: it
verifies already-materialized volumes on pods it is pointed at, with no
knowledge of the serving platform's \storageuri schemes, download code
path, status conditions, or per-namespace policy surface, and it offers no
digest-pinning mode. Section~\ref{sec:integrity} argues for first-party
enforcement inside the platform's own storage initializer---the component
that performs the download the operator never sees---and
Section~\ref{sec:relwork-security} returns to what admission-based tools can and
cannot bind. At the research end, Atlas~\cite{atlas} attests the full ML
lifecycle with transparency logs and trusted hardware; our design is
narrower and deployable today: one enforcement point, standard formats, no
new infrastructure. On the Kubernetes side, admission-based
verifiers---Kyverno image-verification policies~\cite{kyverno} and the
sigstore policy-controller~\cite{policycontroller}---verify image
references present in the pod spec, and today require custom configuration
to extract even image-\emph{volume} references. For artifacts fetched by
the storage initializer (\texttt{s3://}, \texttt{gs://}, \texttt{hf://},
and \ocifetch), the reference appears in the pod spec only as opaque init-%
container arguments: it is not content-addressed, no image verifier
extracts it, and---decisively---no admission-time component can observe the
bytes the initializer will later download. Binding those bytes to a digest
can only happen on the data path. This is the precise gap
Section~\ref{sec:integrity} addresses.

\paragraph{Confidentiality precedent.} KServe recently merged confidential
model serving (PR \#5382): the storage initializer detects JWE-encrypted
artifacts on the same non-OCI URI schemes and decrypts them inside a trusted
execution environment. This established the storage initializer as the
accepted enforcement point for security properties of downloaded models.
Confidentiality answers \emph{who can read} the artifact; the integrity
design in Section~\ref{sec:integrity} is its missing counterpart---\emph{is this
artifact the one the operator signed off on}---on the same code path.

\paragraph{Attacks.} The supply-chain relevance of serve-time artifacts is
not hypothetical: compromise classes demonstrated for software
dependencies~\cite{slsa-solarwinds} transfer directly to models, and recent
surveys of ML supply-chain security~\cite{mlsecops-survey} catalogue
model-replacement and hub-compromise incidents. We defer a full threat model
to Section~\ref{sec:integrity}.

\section{OCI-Native Model Distribution}
\label{sec:distribution}

\subsection{Design space}
\label{sec:design-space}

Once a model is packaged as an OCI image, a Kubernetes serving platform has
three structurally different ways to get its bytes in front of the model
server, summarized in Table~\ref{tab:design-space} alongside the
object-storage baseline. They differ along two axes that matter for the rest
of this paper: \emph{who performs the pull} (the kubelet/container runtime
vs.\ platform code) and \emph{whether the artifact reference appears in the
pod specification}, which determines whether admission-time supply-chain
tooling can see it.

\begin{table*}
\caption{Model delivery paths in KServe. ``Admission-verifiable'' = the
artifact reference appears in the pod spec in a content-addressable form
that admission policy can bind to the bytes the runtime will deliver
(given digest references; Section~\ref{sec:relwork-security} discusses current
tooling maturity). \ocinative and \ocifetch are the schemes contributed by
this work.}
\label{tab:design-space}
\footnotesize
\setlength{\tabcolsep}{4pt}
\begin{tabular}{llllll}
\toprule
Path & Scheme & Pulled by & Node cache & Admission-verifiable & Cluster requirements \\
\midrule
Modelcar sidecar & \ociuri & container runtime & yes & yes & any Kubernetes \\
Image volume & \ocinative & kubelet (KEP-4639) & yes & yes & K8s\textsuperscript{\dag} $\geq$v1.33\textsuperscript{*}, ctrd $\geq$v2.1 / CRI-O $\geq$v1.31 \\
Initializer OCI pull & \ocifetch & storage initializer & no & \textbf{no} & any Kubernetes, registry egress \\
Object storage & \texttt{s3://}, \texttt{gs://}, \texttt{hf://} & storage initializer & no & \textbf{no} & any Kubernetes \\
\bottomrule
\multicolumn{6}{l}{\parbox{0.96\textwidth}{\footnotesize \textsuperscript{*}With the
\texttt{ImageVolume} feature gate on 1.31--1.34 (default-on from 1.35,
stable 1.36); \texttt{subPath} mounts require $\geq$1.33.
\textsuperscript{\dag}K8s = Kubernetes, ctrd = containerd. See Section~\ref{sec:ocinative}.}}
\end{tabular}
\end{table*}

The modelcar pattern (Section~\ref{sec:kserve-bg}) predates both new schemes and
established the image layout convention---model files under
\texttt{/models/} inside the image---that the new paths deliberately
preserve, so a single model image works unmodified across all three OCI
modes. Scheme selection follows the same philosophy: a cluster-wide
\texttt{ociModelMode} default in the platform configuration maps plain
\ociuri URIs to one of the modes, while the explicit \ocinative and
\ocifetch schemes override the default per service. Operators migrate
incrementally; individual teams opt in early.

\subsection{\ocinative: models as image volumes}
\label{sec:ocinative}

The \ocinative implementation (merged upstream; \prnative) maps model
delivery onto the KEP-4639 \texttt{image} volume source. At admission time,
the platform webhook rewrites the pod: the \storageuri becomes a
\texttt{spec.volumes[].image.reference}, and the inference container receives
a read-only mount of that volume at the model path with
\texttt{subPath:\,"models"}, preserving the modelcar layout convention. No
init container is injected---the storage initializer is bypassed
entirely---and no sidecar runs. The kubelet pulls the image through the
container runtime exactly as it pulls the serving image itself, which yields
three properties the download paths lack: node-level caching across pod
restarts and replicas, digest-addressable references with the runtime's
existing content-addressed store, and a reference in the pod spec that
admission controllers can verify \emph{before} the pod is scheduled.

Two design decisions are worth recording. First, \emph{graceful degradation
across the API maturity window}: the image-volume API spans four distinct
capability regimes across Kubernetes releases---alpha and gate-required on
1.31--1.32 (where \texttt{subPath} mounts are additionally rejected), beta
but still gate-required on 1.33--1.34, default-on from 1.35, stable in
1.36. Rather than failing pod creation with an opaque API error, the
controller detects the cluster's regime and surfaces an advisory
\texttt{OciImageVolumeCompatible} condition on the
\texttt{InferenceService}, telling the operator exactly which capability is
missing. Compatibility handling of this kind is unglamorous but was a
precondition for merging: the platform supports clusters spanning several
Kubernetes minor versions.

Second, \emph{bypassing the storage initializer is a feature and a
liability}. It removes an entire copy of the artifact from the startup path
(Section~\ref{sec:eval} quantifies this), but it also means any future
verification logic living in the initializer will never see \ocinative
artifacts. We return to this boundary in Section~\ref{sec:integrity}: for image
volumes, integrity enforcement belongs at admission, where the reference is
visible and mature tooling exists.

\subsection{\ocifetch: portable OCI pull}
\label{sec:ocifetch}

Image volumes require a recent runtime; many production clusters will lack
them for years. \ocifetch (\prfetch, under review at the time of writing)
covers that gap: the storage initializer itself resolves the image reference,
pulls the layers from the registry, and extracts the \texttt{/models/}
tree---no container runtime involvement, no cluster version requirement, and
compatibility with air-gapped registries. The trade-offs invert: the artifact
is re-fetched per pod start (no node cache), and the reference reaches the
pod spec only as an opaque argument to the initializer, not as an image
field any admission verifier extracts.

That last property deserves emphasis because it is easy to miss:
\ocifetch is the \emph{only} OCI delivery path for which no
admission-time verifier can bind the bytes. Kyverno or policy-controller can verify modelcar and
\ocinative references because both appear as image references in the pod
spec and the same runtime that admission vouched for delivers the content;
for \ocifetch, platform code fetches the bytes after every admission
decision is made. Any integrity guarantee for it must be enforced where the
bytes flow---inside the storage initializer---which makes it a natural
early adopter of the verification design in Section~\ref{sec:integrity},
alongside the object-storage schemes it shares that code path with.

\section{Serving-Time Integrity for Non-OCI Storage}
\label{sec:integrity}

The OCI paths of Section~\ref{sec:distribution} inherit the container ecosystem's
verification story. Everything else does not---and ``everything else'' is
the prevailing deployment pattern, as the platform's documentation and
defaults reflect: models referenced as loose objects on \texttt{s3://} and
\texttt{gs://} buckets or \texttt{hf://} hub revisions.
For these, the storage initializer downloads whatever bytes the remote
endpoint returns. This section summarizes the verification design we have
proposed to the KServe community (\rfcint); it is a design under review,
not merged work, and we present it as such.

\subsection{Threat model}

We assume the \texttt{InferenceService} specification and cluster policy are
trusted---an attacker who can rewrite the spec defeats any artifact-level
mechanism and is admission/RBAC territory---while the remote storage
endpoint and the network path are not. We additionally trust the Kubernetes
control plane, the node, the container runtime, and every container that
shares the inference pod: verification happens once, at initialization, and
the verified bytes then sit on a volume shared within the pod, so a
malicious co-located sidecar or an operator with \texttt{pods/exec} could
modify them after the check (a time-of-check/time-of-use window). The
design mitigates but does not close this window---the model volume is
mounted read-only into the server container, and co-located containers are
the cluster policy's problem by assumption; continuous re-validation, as
prototyped by the model-validation-operator's sidecar mode~\cite{mvo}, is
future work. Note also that init containers do not re-run when only the
main container restarts, so cached contents are re-consumed unverified
within the pod's lifetime---another reason verification state must be keyed
on content digest, never on URI (Section~\ref{sec:integrity-design}).

Under that model, the two mechanisms serve different purposes. \emph{Digest pinning}
(an expected digest declared in the spec, checked against the downloaded
bytes, failing closed) defends against artifact mutation, stale or
re-pointed references, corruption, and compromised storage. Because a model
is typically a \emph{directory}, the pin is defined over the artifact's
file-level hash manifest---in the limit, the digest of its OMS
statement---not over an ill-defined concatenation of bytes.
\emph{Signature verification} is the stronger provenance mechanism: it
defends against artifact replacement even when the storage location and any
metadata beside the artifact are attacker-controlled, by requiring the
bytes to verify against a signature from an allowed identity or key. Alone,
however, it verifies \emph{provenance, not intent}: any artifact ever
signed by the allowed identity passes, so an attacker controlling storage
can substitute an older signed revision (a rollback/freeze attack, in The Update Framework (TUF)
terms~\cite{tuf}) or a different model signed by the same release identity.
The policy therefore binds the expected statement subject---model name and,
where available, version---in addition to the signer identity, and we track
freshness requirements as an open question.

\subsection{Design}
\label{sec:integrity-design}

The proposal deliberately invents no cryptography and no format. Signatures
are OpenSSF Model Signing bundles~\cite{modelsigning}---Dead Simple Signing Envelope (DSSE) payloads over
in-toto statements with file-level hashes, supporting keyless (OpenID Connect (OIDC) identity
via Fulcio/Rekor) and key- or certificate-based flows; air-gapped clusters
use the key/certificate flows against a private trust configuration, since
keyless verification requires reaching the public transparency
infrastructure. The platform \emph{consumes} the emerging standard; it does
not compete with it.

Verification runs inside the storage initializer, after download and before
handoff to the model server---the same stage where the merged
confidential-serving support performs TEE-based decryption. The symmetry
is deliberate: confidentiality and integrity are the two halves of one
model-security story on one code path. An optional \texttt{verification} block on
the model spec carries the expected digest and/or signature policy
(bundle location, pinned identity and issuer, or key reference). Failure
behavior is policy, with deliberately asymmetric semantics. In
\texttt{enforce} mode the init container exits non-zero and the pod never
starts serving---and it fails closed on \emph{every} verifier error, not
only on mismatch: a missing bundle, an unreachable trust root, or a crashed
verifier must block startup, because an attacker who controls storage can
trivially delete the signature beside the artifact. \texttt{warn} mode
starts the pod and surfaces the result as an \texttt{InferenceService}
status condition, an event, and a metric; it is a time-boxed migration
state for existing fleets and provides no integrity guarantee---under this
threat model it must not be mistaken for a security control. Policy
resolves per namespace over a cluster-wide default so a platform team can
ratchet enforcement one tenant at a time, with the namespace override
settable only through cluster-scoped RBAC (otherwise a careless tenant
silently downgrades itself). The same discipline applies to the
platform's own delivery pipeline: while upstreaming this work we fixed two
races in the project's CI install path that could validate against a
half-started control plane (\prcia, \prcib)---a policy feature
that silently fails open in CI is worse than no policy feature at all.
Enforce mode also hands a storage-level attacker a startup
denial-of-service; we consider that the correct trade-off and say so
explicitly.

\subsection{Non-goals and boundaries}

Three exclusions keep the design small and are themselves findings about
where enforcement belongs. Modelcar and \ocinative are out of scope: their
references are admission-verifiable (Table~\ref{tab:design-space}), the
runtime that admission vouched for delivers their content, and
image-signature tooling covers them in principle today, and in practice
once verifiers extract \texttt{volumes[].image} references---a
configuration gap, not an architectural one; duplicating that check inside the
platform would create two half-trusted enforcement points. \ocifetch is
architecturally in scope---it transits the initializer and nothing else can
see it---but is deferred to a follow-up to keep the initial surface small.
Packaging-side workflows (ModelPack, KitOps ModelKits) are complementary:
they can \emph{produce} signed artifacts, which this design then
\emph{enforces} at serve time---an enforcement path that also covers the
majority of models that are never repackaged at all.

Open questions tracked in the RFC include the interaction of signature
subjects with encrypted artifacts (verify ciphertext vs.\ verify plaintext:
different guarantees, different signing pipelines), digest pinning against
mutable \texttt{hf://} revisions, freshness requirements against rollback
(Section~\ref{sec:integrity}), verification caching across pod restarts (keyed on
content digest, never URI), and whether hashing should stream during
download rather than run as a second pass. The last question is empirical,
and Section~\ref{sec:eval} answers it.

\section{Evaluation}
\label{sec:eval}

We evaluate two questions: (1)~how do the delivery paths of
Table~\ref{tab:design-space} compare on startup latency and storage cost as
artifact size grows, and (2)~is streaming hash verification cheap enough to
make the Section~\ref{sec:integrity} design's \texttt{enforce} mode practical on
the download path? To our knowledge this is the first systematic,
controlled comparison of model delivery paths within a Kubernetes serving
platform; the container-image distribution literature
(Section~\ref{sec:background}) measures a different object.

\subsection{Methodology}
\label{sec:methodology}

All experiments run on a dedicated cloud VM (AWS \texttt{m6i.2xlarge}:
8~vCPU Intel Xeon Platinum 8375C, 32~GB RAM, us-east-1) with a 1~TB gp3
EBS volume at its baseline provisioning (3{,}000 IOPS, 125~MiB/s), Ubuntu
24.04 (kernel 6.17), Docker 29.1.3, and a single-node \texttt{kind}
v0.32.0 cluster (Kubernetes v1.36.1, containerd 2.3.1, image volumes
enabled), with a colocated OCI registry and in-cluster MinIO as the S3
endpoint---eliminating WAN variance so that differences reflect the
delivery architecture rather than network variability. The provisioned disk
throughput is a ceiling shared identically by every path; absolute times
scale with it, and the object of study is the relative behavior of the
paths on equal hardware. KServe is built from source at upstream commit
\texttt{dbf6cfe}. Artifacts of 2, 14, and 140~GB---sized to the fp16
weights of representative 1B-, 7B-, and 70B-parameter models---combine a
real (tiny) scikit-learn model---so that readiness reflects a genuine
server load---with random ballast, packaged identically as an OCI image (single
layer, \texttt{/models/} layout) and as loose objects in a bucket. (The
measured quantity is the delivery path, which depends on artifact bytes,
not parameter count; the tiers name the model classes those byte sizes
correspond to.) Random ballast is representative, not a shortcut:
production fp16/bf16 weights are already near-incompressible, so
compression-based transfer optimizations help neither. Each cell reports
the median over 10 runs (5 for the I/O-bound 140~GB tier). Interquartile
ranges are within 7\% of the median everywhere except the S3 140~GB
cells (14\% cold, 35\% warm), a spread consistent with the storage-initializer
restarts we observed under memory pressure (Section~\ref{sec:discussion}),
which inflate upper quartiles but not medians; individual outlier runs
of up to $+84\%$ of the cell median occur across the matrix (full
per-run data in the released artifact); on the warm \ocinative{} cells the
absolute IQR is below 0.8\,s.
Phase timings are wall clock with 0.2\,s polling resolution (the
integrity microbenchmark of Section~\ref{sec:eval-integrity} uses its own
in-process timers). \emph{Cold} removes the model image from the node's
containerd store before the run; \emph{warm} starts with it cached. For
the S3 path the ``warm'' condition re-downloads the artifact---there is
no node-level cache on this path, and that asymmetry is precisely the
architectural property the OCI paths exist to change, so we report it as
data rather than omitting it. KServe's optional \texttt{LocalModelCache}
subsystem can pre-warm object-storage models onto dedicated node groups;
it requires pre-warming controllers and node-group configuration
orthogonal to the delivery-path taxonomy measured here, and we scope it
out. The harness and raw data are available at the repository cited in
Section~\ref{sec:intro}.

\subsection{Time to first prediction}
\label{sec:eval-ttfp}

\begin{table}
\caption{Time to first prediction, median seconds. Warm = model image
cached on the node (for S3: re-download, no cache exists).}
\label{tab:ttfp}
\centering
\begin{tabular}{llrrr}
\toprule
 & & \multicolumn{3}{c}{Artifact size} \\
\cmidrule(lr){3-5}
Path & Mode & 2\,GB & 14\,GB & 140\,GB \\
\midrule
\ocinative & cold & 38.6 & 443.4 & 4{,}585.5 \\
           & warm & \textbf{11.5} & \textbf{11.4} & \textbf{11.7} \\
modelcar   & cold & 59.6 & 363.3 & 4{,}603.0 \\
           & warm & 15.4 & 15.2 & 15.8 \\
\texttt{s3://} & cold & 39.6 & 137.8 & 2{,}318.8 \\
           & warm & 30.0 & 128.4 & 2{,}441.6 \\
\bottomrule
\end{tabular}
\end{table}

\begin{figure}
\centering
\includegraphics[width=\columnwidth]{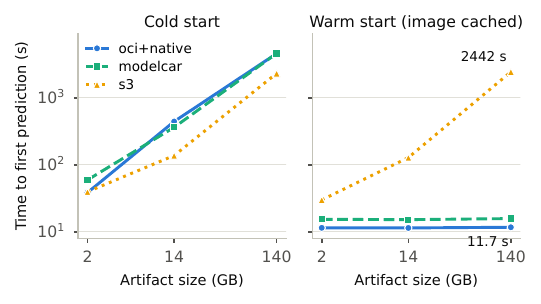}
\caption{Time to first prediction vs.\ artifact size (log--log,
medians). Warm starts on the OCI paths are size-independent; the S3 path
re-downloads on every pod start.}
\label{fig:latency}
\end{figure}

Table~\ref{tab:ttfp} and Fig.~\ref{fig:latency} give the headline result.
\textbf{Warm starts on the OCI paths are size-independent.} With the
model image in the node's containerd store, \ocinative reaches first
prediction in 11.4--11.7\,s whether the artifact is 2~GB or 140~GB;
modelcar tracks it at 15.2--15.8\,s, the constant $\sim$4\,s gap being
sidecar lifecycle overhead. The S3 path has no node cache to warm: every
pod start re-downloads, so its ``warm'' number grows linearly from
30.0\,s to 2{,}441.6\,s. At the 70B-class tier, adding a replica on a
node that has served the model before costs \textbf{11.7\,s with
\ocinative versus 40.7 minutes over object storage---a $208\times$
difference} ($155\times$ for modelcar). This is the number that governs
autoscaling: replicas added under load are overwhelmingly scheduled onto
nodes that have already pulled the image.

\begin{figure}
\centering
\includegraphics[width=\columnwidth]{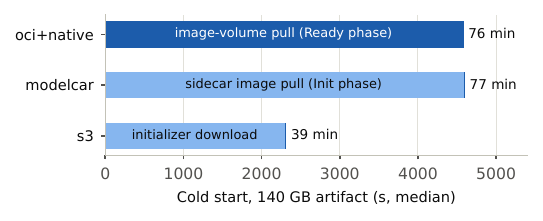}
\caption{Where cold-start time goes at the 140~GB tier (medians). The
OCI paths spend it in the containerd pull (blob write + snapshot
unpack); the S3 path in the initializer's single-pass download.}
\label{fig:phases}
\end{figure}

The cold column tells the honest other half. \textbf{First-pull cost on
the OCI paths is roughly $2\times$ a plain download at the 140~GB tier}
(4{,}585.5 and 4{,}603.0\,s versus 2{,}318.8\,s), and 2.6--3.2$\times$
at 14~GB, where the S3 path additionally benefits from the page-cache
effect discussed below. The mechanism is visible in the phase breakdown
(Fig.~\ref{fig:phases}) and in the kubelet-reported pull durations, which
account for essentially the entire gap: a containerd image pull writes
the layer blob to the content store and then unpacks it into a snapshot
---two passes over the same disk---whereas the storage initializer
streams the object into the pod volume once. (At the 14~GB tier the
initializer path additionally benefits from the colocated MinIO's page
cache; at 140~GB, four times RAM, that advantage disappears and the
$2\times$ ratio is the durable one.) At 2~GB the ordering inverts
---\ocinative (38.6\,s) edges out S3 (39.6\,s) and modelcar pays its
sidecar overhead (59.6\,s)---because fixed per-pod costs dominate
transfer time. The practical rule for operators follows directly:
\emph{the OCI paths win wherever a node serves more than one replica of
a model over its lifetime, and lose only the very first pull}---and the
first pull is exactly what lazy-pulling and compression research on
container images~\cite{slacker} knows how to attack (Section~\ref{sec:discussion}).

\subsection{Storage overhead}

Across the OCI paths, a cold start grows the node's containerd store by
approximately the artifact size (within 5\% at the 14 and 140~GB tiers;
the 2~GB tier additionally carries $\sim$0.7~GB of fixed image overhead),
retained across pod restarts---that retention \emph{is} the warm-start
cache. The S3 path leaves no persistent node state, but materializes a
full transient copy of the artifact in the pod's \texttt{emptyDir} on
\emph{every} start: at the 140~GB tier each concurrent replica requires
140~GB of free node disk in addition to any cached images, and the copy
is re-paid on every restart. Neither path is free; they differ in
whether the space buys anything (a reusable cache) or is pure per-replica
overhead.

\subsection{Cost of integrity verification}
\label{sec:eval-integrity}

\begin{table}
\caption{SHA-256 verification cost relative to plain download
(medians; local object store). The 2\,GB tier is sub-4-second and
page-cache-dominated; conclusions rest on the larger tiers.}
\label{tab:integrity}
\centering
\begin{tabular}{lrrr}
\toprule
 & 2\,GB & 14\,GB & 140\,GB \\
\midrule
Plain download (s) & 1.4 & 190.3 & 2{,}220.4 \\
Streaming hash (s) & 2.3 & 190.4 & 2{,}219.8 \\
\quad overhead & --- & +0.05\% & \textbf{$<$0.1\%} \\
Post-download hash (s) & 3.3 & 203.6 & 3{,}406.8 \\
\quad overhead & --- & +7.0\% & \textbf{+53.4\%} \\
\bottomrule
\end{tabular}
\end{table}

Table~\ref{tab:integrity} answers the feasibility question
Section~\ref{sec:integrity} left open. Hashing the artifact \emph{while} it
downloads is free at any scale that matters: within $\pm$0.1\% of the
plain download at both 14 and 140~GB---SHA-256 throughput on one
core far exceeds the disk- and network-bound download rate, so the hash
hides entirely inside the transfer. Verifying \emph{after} download, as
a separate pass, costs +7.0\% at 14~GB and +53.4\% at 140~GB: the
re-read no longer fits the page cache and becomes a second full pass
over the disk. (The 2~GB tier is sub-4-second and page-cache-dominated---its
download-time variance across variants exceeds the hashing cost being
measured; we report it for completeness but draw conclusions from the
larger tiers.)
The design consequence for \rfcint{} is unambiguous for the
digest-pinning half of the design: verification in \texttt{enforce}
mode should stream during download, where its cost is measurement
noise, and must not be implemented as a post-download pass, where it
adds half again to LLM-scale delivery time. The signature-verification
half adds cryptographic and, in keyless mode, network costs
(Fulcio/Rekor round trips) that are independent of artifact size and
remain to be measured.

\section{Discussion and Lessons}
\label{sec:discussion}

\paragraph{Incrementalism as a security strategy.} None of this work landed
as a big-bang redesign. The distribution schemes arrived as steps of a
community umbrella plan (issue \#4083) whose earlier step (modelcar) had
already normalized OCI models; the integrity proposal explicitly mirrors a
merged confidentiality feature on the same code path. In a large CNCF
project, we found that the fastest route to a stronger supply-chain posture
is to identify the enforcement point the project has \emph{already accepted}
(here, the storage initializer) and extend its guarantees, rather than to
introduce a new privileged component. The same logic dictated what
\emph{not} to build: admission-visible paths stay with admission tooling.

\paragraph{The boundary is the finding.} The most consequential fact in
Table~\ref{tab:design-space} is a negative one: for two of the four
delivery paths, no admission-time component can bind the deployed reference
to the bytes that arrive---not because the reference is hidden (it sits in
the pod spec as initializer arguments), but because nothing at admission
time observes the download. Admission policy can go surprisingly far with
external lookups---resolve a mutable URI, consult a signature service, even
mutate the spec to pin a digest---but resolution at admission and download
at pod start are separated in time, and a mutable reference can change
between them; only a data-path component closes that window. As OCI
packaging pushes models toward image volumes it is tempting to declare the
problem solved at admission; the prevalence of initializer-path deployments
says otherwise. Platform designers should treat ``which component can bind
the bytes'' as the primary question when placing verification, not ``which
component has the best tooling today.''

\paragraph{Initializer paths carry hidden operational cost.} Beyond
latency, running the download inside a pod-level init container means the
download's resource envelope is the operator's problem: at the 140~GB
tier the storage initializer exceeded the memory limits our deployment
had provisioned (sized, like the platform defaults, for classic model
sizes) and was OOM-killed until its limit was raised $4\times$. The
kubelet pull path has no such knob---the runtime manages its own
resources. This is a second, quieter argument for kubelet-managed
delivery at LLM scale, and an operational checklist item for anyone
staying on the initializer path.

\paragraph{Limitations and future work.} Our evaluation is single-node and
LAN-local by design; multi-node image distribution, registry-side rate
limits, and layer-parallel pulls (our artifacts are single-layer) remain
unmeasured, and absolute times are bounded by the provisioned disk
throughput shared by all paths. Our time-to-first-prediction measures
\emph{artifact delivery}: the serving payload is a small classical
model, so GPU inference-engine initialization (weight loading,
tensor-parallel setup), which is additive and delivery-independent for
LLM servers, is out of scope. The colocated object store's page cache
flatters the S3 path at the 14~GB tier (Section~\ref{sec:eval-ttfp}); the
140~GB tier is cache-free and carries the durable ratios. The
$2\times$ cold-pull penalty is not fundamental: zstd layers,
lazy-pulling (eStargz/SOCI-style~\cite{slacker}), and multi-layer
parallelism all attack the blob-write-then-unpack double pass and are
natural follow-ups on the \ocinative path. The integrity design is under community review and may change;
its signature-verification stage depends on the OpenSSF model-signing
format's evolution. Digest pinning does not yet handle mutable \texttt{hf://} revisions
well. \ocifetch verification---the path that needs it most---is specified but
deferred to a follow-up.

\section{Conclusion}
\label{sec:conclusion}

Model artifacts lag container images in supply-chain treatment: mutable
references, unverified downloads, weak policy hooks. This paper reports an
effort to close that gap in a widely deployed CNCF serving platform from
both ends. On the distribution end, the merged \ocinative scheme gives
models kubelet-managed delivery with digest semantics and
admission-verifiable references via Kubernetes image volumes, and \ocifetch
(under review) extends OCI delivery to clusters that cannot use them; our
measurements provide, to our knowledge, the first systematic basis for
choosing among a serving platform's delivery paths. On the verification
end, a serving-time integrity design proposed to the community---digest
pinning and OpenSSF model-signing enforcement at the storage
initializer---covers the object-storage models that no admission-time
verifier can bind, at a streaming-verification cost Section~\ref{sec:eval}
quantifies. The through-line is a simple design rule: put enforcement where
the reference is verifiable when you can, and where the bytes flow when you
must.

\section*{Acknowledgment}

The author thanks the KServe maintainers and reviewers whose feedback
shaped the upstream implementations discussed here. This work received
no external funding. \emph{Disclosure:} the author is the implementer
of the \ocinative and \ocifetch schemes and the author of \rfcint{};
the benchmark harness, raw data, and analysis scripts are released
precisely so that these results can be independently reproduced. In
accordance with IEEE policy on artificial-intelligence tools, the
author discloses that AI-based assistants were used in developing the
benchmark tooling and analysis scripts and in editing the manuscript;
the study design, upstream contributions, validation of all results,
and responsibility for the content rest with the author.


\begin{IEEEbiographynophoto}{Georgii Kliukovkin}
is a software engineer specializing in the efficiency of large-scale
retrieval, ranking, and model-serving systems, and an open-source
contributor to CNCF projects, including KServe, where he authored the
platform's OCI-native model delivery paths. His interests include
machine-learning serving infrastructure, model delivery, and software
supply-chain integrity.
\end{IEEEbiographynophoto}

\end{document}